\documentclass[submission,copyright,creativecommons,leqno]{eptcs}
\providecommand{\event}{NCL 2022} 
\usepackage{breakurl}             
\usepackage{underscore}           

\usepackage[utf8]{inputenc}

\usepackage{bussproofs} 
\usepackage{amssymb} 
\usepackage{amsmath}
\usepackage{mathrsfs}
\usepackage{graphicx}
\newtheorem{theorem}{Theorem}[section]
\author{
	Ji\v{r}{\'\i} Raclavsk\'{y}
	\institute{Department of Philosophy\\
		Masaryk University\\
		Brno, Czech Republic}
	\email{raclavsky@phil.muni.cz}
}
\title{Puzzles of Existential Generalisation\\ from Type-theoretic Perspective}

\def\titlerunning{Puzzles of Existential Generalisation}
\def\authorrunning{J. Raclavsk\'{y}}
\begin{document}
\maketitle

\begin{abstract}
The present paper addresses several puzzles related to the Rule of Existential Generalization, (EG). In solution to these puzzles from the viewpoint of simple type theory, I distinguish (EG) from a modified Rule of Existential Quantifier Introduction which is derivable from (EG). Both these rules are often confused and both are considered as primitive but I show that (EG) itself is derivable from the proper Rule of Existential Quantifier Introduction. Moreover, the latter rule must be primitive in logical systems that treat both total and partial functions, for the universal and the existential quantifiers are not interdefinable in them. An appropriate natural deduction for such a system is deployed. The present logical system is simpler than the system recently proposed and applied by the present author. It utilises an adequate definition of substitution which is capable of  handling not only a higher-order quantification, but also (hyper)intensional contexts.
\end{abstract}


\section{Introduction}
\label{intro}

In literature on \textit{natural deduction} (\textit{ND}), even in the seminal book by Prawitz (\cite[p. 65]{prawitz2006}) one finds two distinct formulations of the \textit{Rule of Existential Generalisation}, (\textit{EG}). Let $\varphi_{(t/x)}$ be the formula $\varphi$ in which all free occurrences of the variable $x$ are substituted by the term $t$; let $F$ be a predicate and $\exists$ be the symbol known as the existential quantifier.
\smallskip

\begin{minipage}{.5\textwidth}
\begin{prooftree}
\AxiomC{$\varphi_{(t/x)}$}
\RightLabel{(EG$_1$)}
\UnaryInfC{$\exists x \varphi(x)$}
\end{prooftree}
\end{minipage}
\begin{minipage}{.4\textwidth}
\begin{prooftree}
\AxiomC{$F(t)$}
\RightLabel{(EG$_2$)}
\UnaryInfC{$\exists x F(x)$}
\end{prooftree}
\end{minipage}
$\newline$

\noindent
Indexing of ``(EG)'' is only done here for convenience. In literature, ``($\exists$-I)'' is often used instead of ``(EG)''.

The ambiguity presents one of several \textit{puzzles} related to (EG) which are addressed in the present paper. Here is their list:

\begin{enumerate}
\setlength\itemsep{.1em}

\item
Having rules for $\forall$, do we also need separate (independent) rules for $\exists$?

\item
Why isn't (EG) called the Rule of Existential Quantifier Introduction, ($\exists$-I)?

\item
Which of (EG$_1$) and (EG$_2$) is the right rule (EG)?

\item
Is (EG$_2$) derivable from (EG$_1$) (or vice versa)?

\item
Why the metalinguistic notation of substitution, $(t/x)$?

\item
Are there problems with quantifying in (hyper)intensional contexts?
\end{enumerate}

\textit{Aims and structure of the paper}.
In the next section, Sec. \ref{solutions}, 
I formulate \textit{solutions} to these puzzles within a type-theoretic approach to logic -- whose foundational example is 
 \textit{simple type theory} (\textit{STT}) by Church \cite{church1940} (see Benzm\"{u}ller and Andrews \cite{benzmuller-andrews2019-sep-ctt} for further references).
Some of these solutions sound familiar, or they are easily derivable from certain assumptions. 
But some of them require more elaborate demonstrations within an appropriate logical framework, which gives rise to the corresponding sections of the present paper.
Needless to add that though the type-theoretic perspective yields a certain reasonable explanation of $\exists$ and rules involving $\exists$, it employs several ideas peculiar to STT which need not be shared by all logical approaches.

In Sec. \ref{logicndtt}, 
I offer an appropriate natural deduction system for \textit{partial type logic} (i.e. a \textit{higher-order logic}) called $\mathsf{TT^*}$ within which I frame some my results. 
$\mathsf{TT^*}$ handles both function-as-mappings and functions-as-computations, which both may not deliver a value for a certain argument.
I indicate whether a given result can already be obtained in the \textit{first-order logic} (\textit{FOL}), or STT of which $\mathsf{TT^*}$ is an extension. In Sec. \ref{egformulationderivation}, 
I show that a proper rule (EG) is derivable from the Rule of Existential Quantifier Introduction, ($\exists$-I), while a modified Rule of Existential Quantifier Introduction, ($\exists$-I$^\eta$), is derivable from (EG). This documents the non-primitiveness of both (EG) and ($\exists$-I$^\eta$) (which corresponds to EG$_1$). In Sec. \ref{syntaxsemanticstt}, 
I also provide a definition of the substitution function and the operator $(t/x)$ which enable us to satisfactorily quantify into both intensional and hyperintensional contexts, Sec. \ref{quantificationinto}. 
Finally, Sec. \ref{conclusion} 
offers a brief conclusion. 

The present paper complements my extensive paper \cite{raclavsky2021-eg} which is currently in print. 
In Sec. \ref{egformulationderivation}, I adjust from it and substantially modify the formulation of (EG) and I provide 
its proof; I adjust the first proof of ($\exists$-I$^\eta$) and borrow several ideas for Sec. \ref{quantificationinto}. However, the present paper
offers an essential revision of the main results of \cite{raclavsky2021-eg}, for it completely removes from $\mathsf{TT^*}$ the (so-called) constructions of the form $\lfloor\!\!\lfloor C \rfloor \!\!\rfloor _\tau$ (so-called immersions), disallowing thus substitution into them. The `objectual reading' of  $C_{(D/x)}$ occurring in (EG) is now secured by constructions which are applications of the function $\lfloor\!\!\lfloor \cdot \rfloor \!\!\rfloor _\tau$ (the choice of similar notation is intentional) which is described below.

\section{Solutions to the puzzles}
\label{solutions}

The present section offers solutions to the puzzles listed in Sec. \ref{intro}, also stating facts concerning $\exists$ and (EG).
(The present section sometimes employs elements of STT, e.g. the notion of type; in Sec. \ref{logicndtt}, we present exact definitions of these notions, though they are adjusted there for  $\mathsf{TT^*}$.)

\subsection{Solution 1}
\label{solution1}

The quantifiers $\forall$ and $\exists$ are well-known to be interdefinable in FOL and its extensions such as STT. Point
b. shows the notation of STT, in which Church's \cite{church1940} $\Pi$ and $\Sigma$ are replaced by $\forall$ and $\exists$.

\begin{itemize}
\small
\item[a.]
$\exists x F(x) =_{df} \neg \forall x \neg F(x)$
\item[b.]
$\exists (\lambda x. F(x)) =_{df} \neg \forall (\lambda x . \neg F(x))$ 
\end{itemize}

\noindent
Recall that in STT, 
$[\![ \lambda x . F(x) ]\!]^{\mathscr{M},v}$ (where $\mathscr{M}$ is a \textit{model}, $v$ is an \textit{assignment} and $[\![ \cdot  ]\!]^{\mathscr{M},v}$ is an \textit{evaluation function}) is a set and $F$ (and even $\lambda x . F(x) $) is a predicate denoting that set.

However, the interdefinability is lost in \textit{partial TT} (``TT'' abbreviates ``type theory'') such as Tich\'{y}'s \cite{tichy1982-partialtt} or $\mathsf{TT^*}$:

\begin{quote}
\textbf{Fact 1} \quad 
\textit{In partial TT, $\forall$ and $\exists$ are not interdefinable (they are independent symbols).}
\end{quote}

\noindent
For suppose $\varphi$ is non-denoting, i.e. $[\![ \varphi ]\!]^{\mathscr{M},v}=\uparrow$, where $\uparrow$ represents an absence of a value. Then, 
$[\![ \forall (\lambda x . \varphi )]\!]^{\mathscr{M},v}= 
[\![ \forall (\lambda x . \neg \varphi )]\!]^{\mathscr{M},v}$, for 
$[\![ \lambda x . \varphi )]\!]^{\mathscr{M},v} =
[\![ \lambda x . \neg \varphi )]\!]^{\mathscr{M},v} = S$, where 
$S$ is the characteristic function undefined for all arguments.
It is thus futile to mimic FOL's definition a. and equate $\neg \exists (\lambda x . \varphi) $ with 
$\forall (\lambda x . \neg \varphi )$.

As a consequence,

\begin{quote}
\textbf{Fact 2} \quad
\textit{In partial TT, $\forall$ and $\exists$ need independent derivation rules.}
\end{quote}

\noindent
 Another consequence is that Henkin's \cite{henkin1963} popular (cf. e.g. \cite{andrews1986,benzmuller-andrews2019-sep-ctt}) reduction of all  logical constants to the equality relations denoted by $\mathit{Q}_{o\tau\tau}$ is lost. Cf. e.g. (using typing \`a Church and many of his abbreviations, notation adjusted)
 $\forall x_\tau F_{o\tau}(x_\tau) =_{df} \
\mathit{Q}_{o(o\tau)(o\tau)} (F_{o\tau},\lambda x_\tau. \mathtt{T}_o)$ and $\varphi_o\land  \psi_o =_{df} \lambda g_{ooo} (g_{ooo} (\mathtt{T}_o,\mathtt{T}_o))= \lambda g_{ooo} (g_{ooo}  (\varphi_o, \psi_o))$, where 
$\mathtt{T}_o =_{df} \mathit{Q}_{ooo}= \mathit{Q}_{ooo}$ 
and $\varphi_o = \psi_o 
=_{df} \mathit{Q}_{ooo} (\varphi_o, \psi_o)$.

\subsection{Solution 2}
\label{solution2}

One of the reasons why (EG) is not called ``the Rule of Existential Quantifier Introduction'' is that the term ``[the Rule of] Existential Generalization'' occurs in Quine's influential article \cite{quine1943}.

But the difference is not merely terminological. We may find that the Rule of Existential Quantifier Introduction and the Rule of Existential Generalization are indeed different.
However, it cannot be ascertained in FOL and its straightforward extensions such as second-order logic, for in them 
$\exists$ of $\exists x \varphi$ is conceived as a certain `\textit{binder}' of variables. But in STT we have a distinct `philosophy' of $\exists$:

\begin{quote}
\textbf{Fact 3} \qquad
\textit{In STT and partial TT, the symbol $\exists$ of $\exists (\lambda x . \varphi)$ is a predicate.}
\end{quote}

\noindent
(From now on the classical $\exists$ as `binder' is never employed.)
The idea that $\exists$ is a \textit{predicate of `non-emptiness'} 
-- alternatively, that it denotes a property of sets `being non-empty' --
goes back to Frege (e.g. \cite{frege1884,frege1951,frege1979-existence}) and seems to be adopted by his great follower Church \cite{church1940}.\footnote{More 
precisely, Church's \cite{church1940} $\exists x_\tau  \varphi_o$ (see p. 58, notation adjusted) stands for $\neg_{oo} \forall x_\tau  \neg \varphi_o$, where
$\forall x_\tau  \phi_o$ stands for $\Pi_{o(o\tau)} (\lambda x_\tau  \phi_o)$. Though Church did not provide an explicit interpretation of $\Pi_{o(o\tau)}$, 
he determined its meaning by the rule of generalization (``from $F_{o\tau } (x_\tau)$ infer 
$\Pi_{o(o\tau)} (F_{o\tau} (x_\tau))$''; see p. 60), which may be read as $\Pi$-introduction rule, 
and the axiom 
$\Pi_{o(o\tau)} (F_{o\tau})
\to 
F_{o\tau} (x_\tau)
$ (see p. 61)
which may be read as $\Pi$-elimination rule. Consequently, his 
$\exists x_\tau \varphi_o$ receives the meaning similar to Frege's. 
But since Church \cite{church1940} allowed arbitrary interpretation of types such as $\iota$ (see p. 57), we may understand him as refraining from the ontological anchorage of Frege's original proposal.
}

Let us state a definition of $\exists$, i.e. a stipulation of the intended meaning of the symbol $\exists$. As mentioned above, in STT without partial functions, $\exists$ is definable using $\forall$, which in turn is definable using $=$. In partial TT, however, $\exists$ must be taken as primitive and its meaning should be stated independently of $\forall$ (possibly via introduction and elimination derivation rules). The following definition of $\exists$ presents the greatest assumption of the present paper concerning $\exists$ (and is in obvious conflict with the FOL view that $\exists$ is a `meaningless symbol').

\begin{quote}
\textbf{Definition of $\exists$} \quad
Let $\tau$ be a type\footnote{A \textit{type} $\tau$ is a symbol denoting a non-empty set of entities, e.g. a set of individuals. Cf. below.} and $o$ be the type of truth values ($\mathtt{T}$ --  True, $\mathtt{F}$ -- false). Let $F$ denote a set (i.e. a characteristic function) of type $\tau \mapsto o$.
The operator ``$\exists$'' is a predicate that denotes the \textit{total} function(-as-mapping) of type $(\tau \mapsto o) \mapsto o$ which is specified by

\[
\exists (F)
  \begin{cases}
    \mathtt{T}, & \text{if the set $[\![F]\!]^{\mathscr{M},v}$ is non-empty} \\
    \mathtt{F}, & \text{otherwise} 
  \end{cases}
\]

\end{quote}

The definition holds even for partial TT. 
In it, one total and many partial \textit{empty sets} map $\tau$-objects to $\mathtt{F}$ or to nothing at all (i.e. they are undefined for a certain argument) while $\exists$ is still applicable to them. In other words, 
the formula $\exists (F)$ has a \textit{definite value} even in partial TT.

Given the above STT's definition of $\exists$, the formula $\exists(F)$ is possible in $\mathscr{L}$ for STT and partial TT, from which we immediately obtain:

\begin{quote}
\textbf{Fact 4} \qquad
\textit{In STT and partial TT, 
the Rule of Existential Quantifier Introduction 
can be stated as a primitive rule}
\vspace{-9pt}
\begin{prooftree}
\AxiomC{$F(t)$}
\RightLabel{($\exists$-I)}
\UnaryInfC{$\exists (F)$}
\end{prooftree}
\end{quote}

Note that our claim that ($\exists$-I) is primitive is confirmed by the fact that it combines only two \textit{primitive ideas} of STT:

\begin{itemize}
\item[a.]
($\exists$-I)'s premiss says: $t$ is an \textit{instance} of $F$
\smallskip
\item[b.]
($\exists$-I)'s conclusion says: $F$ is \textit{nonempty} (= has an instance)
\end{itemize}

As a corollary of our above considerations,

\begin{quote}
\textbf{Fact 5} \qquad
\textit{In classical logics such as FOL, $(\exists$-$\mathrm{I})$ cannot be stated in the form 
\begin{minipage}{.075\textwidth}
\begin{prooftree}
\small
\AxiomC{$F(t)$}
\UnaryInfC{$\exists (F)$}
\end{prooftree}
\end{minipage}
for the reason that $\exists (F)$ is not their well-formed formula (as in TT is), since $\exists$ is not treated as a predicate (as in TT is, cf. the above definition of $\exists$).
}
\end{quote}

\subsection{Solution 3}
\label{solution3}

Once we have identified the rule ($\exists$-I), we may again ask whether the proper (EG) is (EG$_1$) or (EG$_2$). 

(EG$_2$) should be excluded for it apparently arises from ($\exists$-I) by $\eta$-conversion, $F \Leftrightarrow \lambda x. F(x)$, as is clearer from its type-theoretic version:

\vspace{-6pt}

\begin{prooftree}
\AxiomC{$F(t)$}
\RightLabel{($\exists$-I$^\eta$)}
\UnaryInfC{$\exists (\lambda x . F(x))$}
\end{prooftree}

\noindent
In the second part of Sec. \ref{egformulationderivation}, we will see that 
an application of $\eta$-conversion in its proof is not straightforward and is also eliminable.

The only remaining candidate for the proper (EG) is thus (EG$_1$), whose STT's version is the following.

\begin{quote}
\textbf{Fact 6} \qquad
\textit{The only proper rule (EG) is }

\vspace{-9pt}

\begin{prooftree}
\AxiomC{$\varphi_{(t/x)}$}
\RightLabel{(EG)}
\UnaryInfC{$\exists (\lambda x .\varphi(x))$}
\end{prooftree} 

\end{quote}

\noindent
I will adjust this and the above rules for $\mathsf{ND_{TT^*}}$ in Sec. \ref{egformulationderivation}.

\subsection{Solution 4}
\label{solution4}

In both STT and partial TT we thus have

\begin{minipage}{.3\textwidth}
\begin{prooftree}
\AxiomC{$F(t)$}
\RightLabel{($\exists$-I)}
\UnaryInfC{$\exists (F)$}
\end{prooftree}
\end{minipage}
\begin{minipage}{.3\textwidth}
\begin{prooftree}
\AxiomC{$\varphi_{(t/x)}$}
\RightLabel{(EG)}
\UnaryInfC{$\exists (\lambda x .\varphi(x))$}
\end{prooftree}
\end{minipage}
\begin{minipage}{.3\textwidth}
\begin{prooftree}
\AxiomC{$F(t)$}
\RightLabel{($\exists$-I$^\eta$)}
\UnaryInfC{$\exists (\lambda x .F(x))$}
\end{prooftree}
\end{minipage}

\medskip
\noindent
In Sec. \ref{egformulationderivation}, I offer proofs of the two following facts which provide our Solution 4:

\begin{quote}
\textbf{Fact 7} \qquad
\textit{In STT and partial TT (EG) is derivable from ($\exists$-I).}
\end{quote}

\begin{quote}
\textbf{Fact 8} \qquad
\textit{In STT and partial TT ($\exists$-I$^\eta$) is derivable from (EG).}
\end{quote}

\subsection{Solution 5}
\label{solution5}

The use of the \textit{substitution function} \textrm{Sub} via the substitution operator ``$(t/x)$'' is symptomatic of (EG).
Thanks to $(t/x)$, (EG)'s premiss says that an object is such and such. The premiss cannot be true without the existence of such an object.

Some authors (cf. e.g. Uzquiano \cite{uzquiano2020-sep}) ignore this substantial feature of (EG). They claim that (EG) allows us to derive from 
the tautology such as $F(x) \to F(x)$ an empirical truth 
$\exists x (F(x) \to F(x) )$, and so the rule is objectionable, for logic alone cannot pose such metaphysical commitments. However, the authors overlooked that (EG) is not directly applicable to $F(x) \to F(x)$. But once
the premiss is adjusted for a possible use of (EG) to 
$(F(x) \to F(x))_{(t/x)}$, the suspicion that empirical truths are  created from logical ones is dispelled.

Note that the proper character of (EG) changes dependently on the logic in which it is embedded because of the \textit{substitution function} \textrm{Sub} encoded in the notation ``$(t/x)$''. To illustrate, in FOL that manipulates only extensions and in which Sub is a metalinguistic device we have a well-understood rule (of course: one only
replaces terms for individual variables into the few forms of formulas). But in STT that manipulates both extensions and intensions,
and perhaps even hyperintensions, as some logics based on $\mathsf{TT^*}$ do, substitution is complicated because of the 
mutual and sometimes tricky interaction of extensions, intensions and hyperintensions. Cf. our treatment (in Sec. \ref{quantificationinto}) of examples listed in Sec. \ref{solution6}, which relies on our definition of Sub in Sec. \ref{substitution}.

Given the fact that ``$(t/x)$'' is a metalinguistic symbol of both FOL and STT (cf. e.g. Church \cite{church1940}),

\begin{quote}
\textbf{Fact 9} \qquad
\textit{The operator $(t/x)$ is not a proper part of logics such as FOL or STT. But it may be a proper part of e.g. Tich\'{y}'s logic  \cite{tichy1988-ffl} and its derivatives such as $\mathsf{TT^*}$ which have `levels' in which they accommodate metalinguistic notions.}
\end{quote}

\subsection{Solution 6}
\label{solution6}

In a number of writings, e.g. \cite{quine1943,quine1956}, Quine and his followers (e.g. Kaplan \cite{kaplan1968}) debated \textit{quantification into}. Quantification into is needed e.g. when one illuminates e.g. a. \textit{de dicto} and b. \textit{de re} readings of belief sentences ``\textsl{$A$ believes that $x$ is an $F$}'' using $\exists$:
a.  $Bel (a, \exists (\lambda x. F(x)))$, b.  $\exists (\lambda x  . Bel (a, F(x)))$.
But a more typical use of quantification into occurs when one applies (EG).

Quine then repeated Frege's seminal observation that one cannot substitute something (via the application of Leibniz's Rule) into \textit{indirect contexts} which are created by verbs such as ``believe''. Quine called them \textit{opaque contexts} and contrasted them with \textit{transparent contexts} in which quantification into is unproblematic. Note that the substantial issue here is substitution,  since quantification into is based on it.

Recently, opaque contexts were split into \textit{intensional} and \textit{hyperintensional contexts} (briefly \textit{I-} and \textit{H-contexts}) and they were contrasted with \textit{extensional contexts} (\textit{E-contexts}). I-contexts cover \textit{modal contexts} (e.g. ``\textsl{it is necessary that ...}''). In both I- and H-contexts, however, only \textit{de dicto readings} of the expressions involved in them make  substitution into them impossible. On their \textit{de re readings}, on the other hand, the intended substitutions are possible as in E-contexts, since the expressions for which we substitute lie outside the scope of the `opacity' operators (or they are only seemingly in its scope). To illustrate, 
``\textsl{$B$ is such that $A$ believes that she [= $B$] is an $F$}'' or ``\textsl{$A$ believes of $B$ that she [= $B$] is an $F$}'' contain ``$B$'' that lies outside the scope of ``\textsl{believes}''.

A \textit{context} $C$ is called \textit{E-, I-}, or \textit{H-context} iff it is the identity of either the extensions, intensions, or hyperintensions of expressions ``$E_1$'' and ``$E_2$'' that is required when we substitute one for another within $C$. 
(Needless to add that the result of substitution must be congruent with the input formula.)
Consequently,

\begin{quote}
\textbf{Fact 10} \qquad
\textit{Quantification into E-, I-, H-contexts using (EG) requires an appropriate management of substitution in the logic appropriate for these contexts.
}
\end{quote}

In Sec. \ref{logicndtt}, I offer such a logic and also define there the substitution function (and thus the operator denoting it) which successfully treats quantification into using (EG). The following quadruple of arguments provide our test examples.
Let $D$ be e.g. ``\textsl{the King of France}'' and $F$ be e.g. ``\textsl{[is a] king}''.
\begin{center}
\begin{minipage}{.4\textwidth}
\begin{prooftree}
\small
\AxiomC{\textsl{$D$ is necessarily an $F$.}}
\LeftLabel{$A^I_1$}
\UnaryInfC{\textsl{There is an individual which is necessarily an $F$.}}
\end{prooftree}
\end{minipage}
\qquad
\begin{minipage}{.4\textwidth}
\begin{prooftree}
\small
\AxiomC{\textsl{Necessarily, $D$ is an $F$.}}
\LeftLabel{$A^I_2$}
\UnaryInfC{\textsl{There is an individual such that }}
\noLine
\UnaryInfC{\textsl{necessarily, he's an $F$.}}
\end{prooftree}
\end{minipage}
\end{center}

\noindent
Obviously, $A^I_1$ involves an unproblematic modality \textit{de re}, while 
$A^I_2$ involves modality \textit{de dicto} in which, on the other hand, substitution is unwarranted. The following two arguments discuss an abortive (improper) computation $3 \div0$.

\begin{center}
\begin{minipage}{.4\textwidth}
\begin{prooftree}
\small
\AxiomC{\textsl{$3{\div}0$ is improper.}}
\LeftLabel{$A^H_1$}
\UnaryInfC{\textsl{There is a computation which is improper.}}
\end{prooftree}
\end{minipage}
\qquad
\begin{minipage}{.4\textwidth}
\begin{prooftree}
\small
\AxiomC{\textsl{$3{\div}0$ is improper.}}
\LeftLabel{$A^H_2$}
\UnaryInfC{\textsl{There is a number $n$ such that}}
\noLine
\UnaryInfC{\textsl{ $3{\div}n$ is improper.}}
\end{prooftree}
\end{minipage}
\end{center}

\noindent
$A^H_1$ involves a `\textit{de re}' H-context, as we may perhaps characterize it, since the indicated quantification into it is unproblematic, while
$A^H_2$ seems to involve `\textit{de dicto}' $H$-context, since 
the indicated quantification into it appears unwarranted.

\section{$\mathsf{TT^*}$ and its natural deduction}
\label{logicndtt}

As indicated above, there are several reasons why to choose STT enriched by partial functions (for more on this choice, see e.g. Farmer \cite{farmer1990}). 
Moreover, to adequately formalize the above examples we need STT which is extended for the purpose of the \textit{fine-grained analysis} of intensional or hyperintensional contexts. 
A suitable logic which meets such requirements is Tich\'{y}'s \cite{tichy1988-ffl} and then e.g. the proposal by Moschovakis \cite{moschovakis2006-meaningandsynonymy}. In \cite{raclavsky2020-book}, Kuchy\v{n}ka's (unpublished) effective modification of Tich\'{y}'s approach to the formalization of language was elaborated. Further modifications led to the development of $\mathsf{TT^*}$ (cf. \cite{raclavsky2021-eg}) which is briefly described below.

The key concept of Tich\'{y}'s neo-Fregean algorithmic approach consists in fine-grained hyperintensions called \textit{constructions}, 
alluding here to geometry where (say) one intersection  can be constructed by infinitely many congruent ways. 
Similarly, the truth value $\texttt{T}$ (True) can be constructed by the application of $\neg$ (negation) to $\texttt{F}$ (False), or much less effectively by $\forall x \forall y \forall z \forall n (( x^n+y^n=z^n)\to (n < 3))$. Constructions are thus not necessarily effective, acyclic algorithmic computations of an object.

\subsection{Language and semantics of $\mathsf{TT^*}$}
\label{syntaxsemanticstt}

Constructions are best representable by $\lambda$-terms; 
I will use the following notation (of which $(,), \lambda \tilde{x}_m., \ulcorner, \urcorner$ are syncategorematic symbols). 
Recall that we use a substantial modification of Tich\'{y}'s logic, $\mathsf{TT^*}$. Let $\tau$ be a type (see below) and let ``$\tilde{E}$'' be short for the string of entities $E_1 ... E_m$; ``$\bar{E}$'' be short for $E_1, ..., E_m$. 
$$
\mathscr{L}_{\mathsf{TT^*}}
\qquad
C := \mathbf{X} \, | \, x \, |  \,  C_0(\bar{C}_m) \, | \,  \lambda \tilde{x}_m . C_0 \, | \,  \ulcorner C_0 \urcorner 
$$
I will also use auxiliary brackets [, ]; boldface of $\mathbf{X}$ will usually be suppressed. 

{\small
\textit{Remark}. The last two terms of $\mathscr{L}_{\mathsf{TT^*}}$ presents our expansion of the (usual) language $\mathscr{L}_{\mathrm{STT}}$. The language $\mathscr{L}_{\mathrm{FOL}}$ differs from $\mathscr{L}_{\mathrm{STT}}$ by having FOL's $\exists$ or $\forall$ instead of $\lambda$ and by restriction to a few particular forms of the other terms, e.g. $\to(\varphi,\psi)$ instead of the much more general $C_0(\bar{C}_m)$.
STT also extends FOL by allowing variables ranging not only over the domain of individuals, but having also additional variables ranging even over domains of various functions (recall that \textit{sets} are identified in STT with characteristic functions). Our formulation of $\mathscr{L}_{\mathsf{TT^*}}$ is schematic, but I will add particular constants (e.g. the logical constants $\neg, \exists$) when needed.
}

The above BNF definition entails the notion of \textit{subconstruction} of a construction. 
An occurrence of $x$ is \textit{bound in} $C$ iff it occurs in $C$ within the scope of $\ulcorner . \urcorner$ or $\lambda \bar{x}_m $ (provided $x$ is one of $\bar{x}_m$), but $x$ is \textit{free in} $C$ iff at least one of its occurrences is not bound in $C$. 

Each construction constructs dependently on \textit{assignment} $v$, we say that it $v$-\textit{constructs} an object. Notation: $v$ that assigns to $x$ the object $\mathrm{D}$ is denoted by ``$v(\mathrm{D}/x)$''. Some constructions, e.g. $\div (3,0)$ (below written rather $3\div 0$), $v$-construct nothing at all (which I will denote by ``$\_$'' as e.g. in \cite{raclavsky2021-eg}), they are called $v$-\textit{improper}. \textit{Partiality} already occurs on the level of functions(-as-graphs): both \textit{total} and \textit{partial} multiargument  \textit{functions} (e.g. $\div$) are manipulated in $\mathsf{TT^*}$;\footnote{Tich\'y \cite{tichy1982-partialtt} disproved Sch\"onfinkel's reduction in the case with partial functions.} a function is called \textit{partial} iff it is undefined for at least one member of its domain (of arguments).

The exact behaviour of constructions was described in \cite{raclavsky2018-existimport,raclavsky2020-book} in terms of \textit{Henkin-style model-theoretic semantics}, see Sec. \ref{sec:appendix} (Appendix), in which the notions of type and model are used (see below).
Here I only state a brief characterisation.
A \textit{constant}-as-construction $\mathbf{X}$ $v$-constructs the object $\mathrm{X}$ (mind the upright font) that is assigned to the term ``$\mathbf{X}$'' by the \textit{interpretation function} $\mathscr{I}$. 
A \textit{variable} $x$ is a construction that $v$-constructs an object assigned to it by $v$. 
Each variable is tied to a particular variability \textit{range}, which is an interpretation of a certain type.
An \textit{application} $C_0(\bar{C}_m)$ $v$-constructs the value (if any) of the function (if any) $v$-constructed by $C_0$ at the $m$-tuple $v$-constructed by  $\bar{C}_m$; otherwise $C_0(\bar{C}_m)$ $v$-constructs nothing, it's $v$-improper.
A $\lambda$-\textit{abstraction} $\lambda \tilde{x}_m . C_0$
$v$-constructs a function 
whose arguments are $m$-tuples consisting of members of the Cartesian product of domains (ranges) corresponding to $\bar{x}_m$ and whose values are $v$-constructed by $C_0$ on the respective $v$ and its appropriate modifications $v'$ which are like $v$ except for what $v$ assigns to $\bar{x}_m$.
An \textit{acquisition} $\ulcorner C_0 \urcorner $ $v$-constructs the entity $C_0$ as such. 

Two constructions are called $v$-\textit{congruent} iff they $v$-construct the same object, or they are both $v$-improper. 

Each term/construction is typed via \textit{type statements} of the form $C/\tau$. An object belonging to type $\tau$ of order $n$ is called an \textit{$n$th-order $\tau$-object}. The full definition of the notions of \textit{type} and \textit{order} can be found in  
\cite{raclavsky2020-book} or \cite{raclavsky2018-existimport}; I only state here its core.

Types $\tau$ are interpreted by \textit{domains} $\mathscr{D}_\tau$ which each is a set of (all) objects of a certain `kind'.
A \textit{frame} $\mathscr{F}$ consists of an indexed family $\{ \mathscr{D}_\tau \}_{\tau \in \mathscr{T}}$ of domains that interpret all types in a set $\mathscr{T}$ of types. A \textit{model} $\mathscr{M}$ is a couple $ \langle \mathscr{F} , \mathscr{I}\rangle $ (in Henkin general models, domains of function types only consist of some, not necessarily all, possible functions of the type in question).

Let $1 \leqslant n  \in \mathbb{N}$. Let our \textit{type base} $\mathscr{B}=\{\iota,o\}$ which I occasionally extend by adding $\nu$ or $\omega$, where\footnote{$\mathscr{B}=\{\iota,o\}$  can easily be made more `topic-neutral' -- as is usual in logic -- by defining truth values as certain entities (see e.g. Henkin \cite{henkin1963} or Andrews \cite{andrews1986}), while
$\iota$'s interpretation $\mathscr{D}_\iota$ may even be left empty. The types such as $\omega$ are peculiar to a specific theory build within the framework of $\mathsf{TT^*}$. Further remark: 
 $x$ is sometimes used as ranging over an unspecified type $\tau$, which is clear from the context surrounding it.
}

\begin{center}
\small
\begin{tabular}{lll}
\textit{Type} $\tau$ & $\mathscr{D}_\tau$ \textit{consists of} & \textit{variables ranging over $\tau$}\\
\hline
$\iota$ & \textit{individuals} & $x, y, ...$ \\

$o$ & \textit{truth values} 
& $o, o', ...$ \\

$\nu$ & \textit{natural numbers} & $n, n', ...$ \\

$\omega$ & \textit{possible worlds} & $w, w', ...$ \\
\end{tabular}
\end{center}

\begin{itemize}
\small
\setlength\itemsep{0.1em}
\item[$\tau^1$]
\textit{$1$st-order types}:
a. all members of $\mathscr{B}$ are $1$\textit{st-order types}, and b. if $\bar{\tau}_m$ and $\tau_0$ are $1$st-order types, $\langle \bar{\tau}_m \rangle \mapsto \tau_0$ is also a $1$\textit{st-order type}, denoted by $\langle \bar{\tau}^1_m \rangle \mapsto \tau^1_0$; $\mathscr{D}_{\langle \bar{\tau}^1_m \rangle \mapsto \tau^1_0}$ consists of total and partial functions from the Cartesian product of $\bar{\mathscr{D}}_{\bar{\tau}^1_m}$ to $\mathscr{D}_{{\tau}^1_0}$. 

\item[$*^n$]
$n$\textit{th-order constructions} are all constructions whose subconstructions $v$-construct objects of $n$th-order types. 

\item[$\tau^{n+1}$]
$(n{+}1)$\textit{st-order types}: a. each $n$th-order type is an $(n+1)$\textit{st-order type}; b. the type $*^n$ is an $(n{+}1)$\textit{st-order type} ($\mathscr{D}_{*^n}$ consists of all $n$th-order constructions), and, c. if $\bar{\tau}_m$ and $\tau_0$ are $n$th-order types, then $\langle \bar{\tau}_m \rangle \mapsto \tau_0$ is also an $(n{+}1)$\textit{st-order type}, denoted by $\langle \bar{\tau}^{n+1}_m \rangle \mapsto \tau^{n+1}_0$;  $\mathscr{D}_{\langle \bar{\tau}^{n+1}_m \rangle \mapsto \tau^{n+1}_0}$ consists of total and partial functions from the Cartesian product of $\bar{\mathscr{D}}_{\bar{\tau}^{n+1}_m}$ to $\mathscr{D}_{\tau^{n+1}_0}$.
\end{itemize}

\noindent
This and the specification of constructions' behaviour entail that each $n$th-order construction consists of 
\textit{$n$th-order subconstructions}, with exceptions of constructions of the form $C:= \ulcorner C_0\urcorner$ 
of which $C_0$ is a \textit{lower-order} (by 1) subconstruction.
A construction $C \in \mathscr{D}_{*^n}$, where $1 \leqslant n$, is called and \textit{$n$th-order construction}.

\textit{Typing statements} are of the following forms:
i. $x/\tau^n$; ii. $\mathbf{X}/\tau^n$; iii.
$C_0(\bar{C}_m)/\tau^n_0$, where $C_1/\tau^n_1; ... ;$
$ C_m/\tau^n_m; C_0/\langle \bar{\tau}^n_m \rangle \mapsto \tau^n_0$;
iv. 
$\lambda \tilde{x}_m. C_0/ \langle \bar{\tau}^n_m \rangle \mapsto\tau^n_0$, where $x_1/\tau^n_1; ...; x^n_m/\tau^n_m; C_0/\tau^n_0$;
v. $\ulcorner X \urcorner/ \tau^n$, where $X/\tau^{n{-}1}$. 
From now on, upper indices in the records of types $\tau^n$ (but not in $*^n$) will be suppressed.

\subsection{Explicit substitution}
\label{substitution}

For my rule (EG) and other rules to be fully specified, the operator $(t/x)$ must be clearly defined.
For that reason I utilize a technique of \textit{explicit substitution}.

In \cite{raclavsky2021-eg}'s version of $\mathsf{TT^*}$, I used a (modification) of Tich\'{y}'s \cite{tichy1988-ffl} special kind of constructions 
$ \lfloor\!\!\lfloor C \rfloor\!\!\rfloor _\tau $ (denoted by him $^2C$) that doubly execute $C$. For example, 
$ \lfloor\!\!\lfloor \ulcorner x \urcorner \rfloor\!\!\rfloor _\tau $ $v$-constructs the same $\tau$-object as $x$. Which is why I allowed substitution in e.g. $ \lfloor\!\!\lfloor \ulcorner x \urcorner \rfloor\!\!\rfloor _\tau $ (Tich\'{y} did not, despite that $x$ is evidently free in it). Recently, Kuchy\v{n}ka (personal communication in spring 2021) revealed inconsistencies resulting from it. Below, I present a way 
how to eliminate $ \lfloor\!\!\lfloor C \rfloor\!\!\rfloor _\tau $ without an undesirable loss of expressive power.

The present version of $\mathsf{TT^*}$ utilizes a partial function 
denoted by $ \lfloor\!\!\lfloor \cdot \rfloor\!\!\rfloor _\tau $. 
The function maps every $n$th-order construction $C$ to the object (if any) $v$-constructed by $C$; the function is thus generally specified by the evaluation function defined in \cite{raclavsky2021-eg} (see Sec. \ref{sec:appendix}), alternatively by derivation rules of $\mathsf{ND_{TT^*}}$.
The symbol for the construction of the function, 
$ \lfloor\!\!\lfloor \cdot \rfloor\!\!\rfloor _\tau / *^n \mapsto \tau^n$, is a part of $\mathscr{L}_\mathsf{TT^*}$ that enables the naming of the function.
For convenience, I will write applications applying that function to some construction $C$ as $\lfloor\!\!\lfloor C\rfloor\!\!\rfloor _\tau$, not in a prefix way.

\begin{theorem}
[$C \cong \lfloor\!\!\lfloor \ulcorner C \urcorner \rfloor\!\!\rfloor _\tau$]
Let $C /\tau^n; \lfloor\!\!\lfloor \cdot \rfloor\!\!\rfloor _\tau / *^n \mapsto \tau^n$. Then,
$C \cong \lfloor\!\!\lfloor \ulcorner C \urcorner\rfloor\!\!\rfloor _\tau.$
\end{theorem}

\textit{Proof}.
$C$ is either a. $v$-proper, or b. $v$-improper. By specification of constructions, $C$ $v$-constructs a. a certain object $\mathrm{C}$ (mind the upright font), or b. nothing at all.
In case a., $\lfloor\!\!\lfloor \ulcorner C \urcorner\rfloor\!\!\rfloor _\tau$ 
$v$-constructs the very same object $\mathrm{C}$, by specification of acquisitions and applications and the specification of the function $\lfloor\!\!\lfloor \cdot \rfloor\!\!\rfloor _\tau$.
Similarly for case b. when $\lfloor\!\!\lfloor \ulcorner C \urcorner\rfloor\!\!\rfloor _\tau$ $v$-constructs nothing at all.
In both cases, the conditions of the Theorem are satisfied.

{\small
\textit{Remark.}
The theorem can be converted into three rules of 
an appropriate natural deduction system for $\mathsf{TT^*}$, roughly of the forms $C \vdash 
\lfloor\!\!\lfloor \ulcorner C \urcorner\rfloor\!\!\rfloor _\tau$ (introduction rule), 
$\lfloor\!\!\lfloor \ulcorner C \urcorner\rfloor\!\!\rfloor _\tau
 \vdash C$ (elimination rule), 
and a special form of elimination rule called instantiation rule which can be found below (cf. $\lfloor\!\!\lfloor \cdot
\rfloor\!\!\rfloor_\tau$-INST).
}

If convenient, I will utilise $$C_{(D/x)}$$
as short for 
$$ \lfloor\!\!\lfloor Sub(\ulcorner D \urcorner, \ulcorner x \urcorner,\ulcorner C \urcorner) \rfloor\!\!\rfloor_\tau.$$

The `operator' $Sub$ (unabbreviated notation: $Sub^n$) mentioned in (the unabbreviated form of) $C_{(D/x)}$
$v$-constructs the function Sub$^{(n)}$  which is of type $\langle *^n,*^n,*^n \rangle \mapsto *^n$.
Sub maps triples of constructions such as $\langle D,x,C\rangle$ to constructions which I also denote by ``$C_{(D/x)}$''. It is important to state that $C_{(D/x)}$ is 
$v(\mathrm{D}/x)$-congruent with $C$, which is captured by the \textit{Compensation} (or \textit{Substitution}) \textit{Principle} (for its full proof, see the relevant part of \cite{raclavsky2020-book}). 
Notation: $C_{(\bar{D}_m/\bar{x}_m)}$ is short for
$C_{(D_1/x_1)...(D_m/x_m)}$.

{\small
\textit{Remark.}
Note that we employ here a higher `level' (by 1) of our language, since the constructions $D,x,C$ are $C_{(D/x)}$'s lower-order subconstructions. 
Which shows that $C_{(D/x)}$ cannot adequately be handled in STT or FOL.
}

My definition of Sub is an unsubstantial modification of Tich\'{y}'s \cite{tichy1982-partialtt} in the style that is more usual in $\lambda$-calculus (see e.g. Hindley and Seldin \cite{hindley-seldin2008}). But it significantly differs from his and other substitution functions since these are only occurring in meta-language; for us, the substitution function and the constructions deploying it are internal objects of the $\mathsf{TT^*}$ framework.

\begin{quote}
\small
\textbf{The substitution function Sub}
Let $C,D,x,B$ are $n$th-order constructions.
Let ``$FV(C)$'' stand for the set of all free variables that are subconstructions of $C$.

\medskip

I. \qquad If the variable $x$ is not free in $C$, then $C_{(D/x)}$ is identical with $C$.

II.  $\!\!\!\!\!$ \qquad
 If the variable $x$ is free in $C$, then

\begin{center}
\small

\begin{tabular}{l|p{60pt}|p{75pt}p{250pt}}

& \textit{If $C$ is ...}& \textit{$C_{(D/x)}$ is ...} & \textit{condition:}\\

\hline
i. & $x$ & $D$ \\

ii. & $B (\bar{B}_m)$ & $B_{(D/x)} (\bar{B}_{m(D/x)})$ \\

iii. & $\lambda y . B$ & $\lambda y .B_{(D/x)}$ & 
\footnotesize $x  \in FV(B)$ and $y \not \in FV(D) $ \\

iv. & $\lambda y . B$ & $[\lambda z . B_{(z/y)}]_{(D/x)}$ & 
\footnotesize 
$x  \in FV(B)$ and $y \in FV(D) $ and $z \not \in FV(B) \cup FV(D) $ \\

\end{tabular}
\end{center}
\end{quote}

{\small
\textit{Remark.} FOL requires (a meta-linguistic version of) Sub only defined by I., II.i-ii and  modified points II.iii-iv (with FOL's $\exists, \forall$ instead of $\lambda$). STT requires Sub defined by I., II.i--iv, 
while impossibility to substitute in $\ulcorner C \urcorner$ is not assumed in point I.
}

\subsection{Natural deduction for $\mathsf{TT^*}$}
\label{naturaldeductiontt}

The \textit{natural deduction} system \textit{in sequent style} for Tich\'{y}an logics exists, see Tich\'{y} \cite{tichy1982-partialtt,tichy1986-indiscernibility} and  Raclavsk\'y (et al.) \cite{raclavsky-kuchynka-pezlar2015, raclavsky2020-book}, where one finds Kuchy\v{n}ka's rules for constructions of the form $\ulcorner C \urcorner $ which are missing in Tich\'{y}'s 
writings.\footnote{For techniques of ND, see e.g. 
Negri et al. \cite{negri-plato-ranta2001}, 
Indrzejczak \cite{indrzejczak2010},
Quieroz et al. \cite{quieroz-oliviera-gabbay2011}.
}
The system is called $\mathsf{ND_{TT^*}}$.

Its rules are made from sequents while sequents are made from  congruence statements of the form 

\begin{center}
$C{:}x$ or $C{:}\mathbf{X}$ or $C{:}\ulcorner C_0 \urcorner$, 
\end{center}
called \textit{matches} $\mathtt{M}$.  Notation: $C{:}\mathbf{x}$ indicates all three variants of matches. In a model $\mathscr{M}$ (similarly below), $v$ \textit{satisfies} $\mathtt{M}$ iff $\mathtt{M}$'s $C$ and $x$ (or $X$ or $\ulcorner C_0 \urcorner$) are $v$-congruent, i.e. they $v$-construct the same object.
Empty matches of the form $C{:}\_$, where $C$ is $v$-improper, are also allowed. Empty matches say that $C$ is $v$-improper, while the other matches say that $C$ is $v$-proper and some even suggest which object $C$ $v$-constructs. The system thus handles `signed formulas', which rather increases our inferential capability. Moreover, matches allow us to escape Blamey's \cite{blamey1986} influential criticism of partial logic, for each match is definitely satisfied or counter-satisfied and so his counter-examples do not apply to $\mathsf{ND_{TT^*}}$.

A \textit{sequent} $\mathtt{S}$ is a couple of the form $$\Gamma \longrightarrow \mathtt{M},$$ 
where $\Gamma$ is a finite set of matches and $\mathtt{M}$ a match.
$\mathtt{S}$ is \textit{valid} iff every $v$ which satisfies all members of $\mathtt{S}$'s $\Gamma$ satisfies also $\mathtt{S}$'s $\mathtt{M}$.
A \textit{rule} $\mathtt{R}$ is a validity preserving operation on sequents of the form 

\begin{center}
\begin{minipage}{.06\textwidth}
\begin{prooftree}
\AxiomC{$\bar{\mathtt{S}}_m$}
\UnaryInfC{$\mathtt{S}$}
\end{prooftree}
\end{minipage}
\end{center}

\noindent
$\mathtt{R}$ is \textit{valid} iff every $v$ that satisfies $\mathtt{R}$'s $\bar{\mathtt{S}}_m$ also satisfies $\mathtt{R}$'s $\mathtt{S}$.
 I write ``$\Gamma, \Delta $'' (where $\Delta$ is another set of matches) instead of ``$\Gamma \cup \Delta $'' and
``$\Gamma , \mathtt{M} $'' instead of ``$\Gamma \cup \{ \mathtt{M} \} $''.
A \textit{derivation} $\mathtt{D}$ is a (finite) list of sequents in which each sequent arises as an application of an inference rule to some preceding sequents or sequents occurring in the set $H$ of sequents given as hypotheses; $\mathtt{D}$ is also called the \textit{proof of} the last \textit{sequent} in the list; we will use a tree presentation of proofs.

In the present derivation of (EG) I will deploy the following selected rules of $\mathsf{ND_{TT^*}}$ (for full list of its rules, see \cite{raclavsky2021-eg}). 
If not stated otherwise, let  
 $x, A, \mathbf{y}, Y, D/\tau; 
 F, f $
 $/\langle \bar{\tau}_m \rangle \mapsto \tau; 
 X_1,x_1/\tau_1; ...; X_m, x_m/\tau_m$. 
 Further, let $o, \mathrm{T} , C /o$ ($\mathrm{T}$ $v$-constructs $\mathtt{T}$).
\textit{Conditions} of the rules 
should include that the variables occurring within the rules are pairwise distinct and are not free in $\Gamma, \mathtt{M}$ and other constructions occurring in the rule (but, of course, $x$ typically occurs freely in $C$ of $C_{(D/x)}$).
Let us assume no mishandling of orders. 

Moreover, we will need to adjust some rules below to manage also \textit{intensions} $I$ (as functions from possible worlds) which are  $v$-constructed by certain constructions $D/\omega \mapsto \tau$; the value (if any) of $I$ at a $v$-value of $w$ is $v$-constructed by $D(w)$; let 
$D_{(k)}$ be $D_{w}$ if $k=1$ but $D$ if $k=0$ (similarly, but the other way around, for types:
let $\tau_{(k)}$ be $\tau$ if $k=1$ but $\omega \mapsto \tau$ if $k=0$). Let $D_{(k)}/ \tau_{(k)}; d/\tau;\lfloor\!\!\lfloor \cdot \rfloor\!\!\rfloor_\tau / *^n \mapsto \tau^n; 
0 \leqslant k \leqslant 1$. 
Let $\mathscr{L}_{\mathsf{TT^*}}$ be further extended by the constant
for $\exists^\tau/ (\tau \mapsto o)\mapsto o$.

\small

\begin{minipage}{0.3\textwidth}
\begin{prooftree}
\AxiomC{}
\RightLabel{(AX)}
\UnaryInfC{$\Gamma, \mathtt{M} \longrightarrow \mathtt{M}$}
\end{prooftree}
\end{minipage}
\begin{minipage}{0.3\textwidth}
\begin{prooftree}
\AxiomC{$\Gamma \longrightarrow \mathtt{M}$}
\RightLabel{(WR)}
\UnaryInfC{$\Gamma, \Delta\longrightarrow \mathtt{M}$}
\end{prooftree}
\end{minipage}
\begin{minipage}{0.3\textwidth}
\begin{prooftree}
\AxiomC{$\Gamma  \longrightarrow F (A) {:} \mathrm{T}$}
\RightLabel{($\exists$-I)}
\UnaryInfC{$\Gamma \longrightarrow \exists^\tau (F) {:} \mathrm{T}$}
\end{prooftree}
\end{minipage}

\medskip

\qquad
\qquad
\qquad
\qquad
\begin{minipage}{0.5\textwidth}
\begin{prooftree}
\AxiomC{$\Gamma \longrightarrow  Y_{(\bar{X}_m/\bar{x}_m)} {:} {\mathbf{y}}$}
\AxiomC{$\Gamma \longrightarrow X_1 {:} \mathbf{x}_1$; ... ; $\Gamma \longrightarrow X_m {:} \mathbf{x}_m$}
\RightLabel{($\beta$-EXP)}
\BinaryInfC{$\Gamma \longrightarrow [\lambda \tilde{x}_m . Y] (\bar{X}_m) {:} {\mathbf{y}}$}
\end{prooftree}
\end{minipage}

\medskip

\begin{minipage}{0.5\textwidth}
\begin{prooftree}
\small
\AxiomC{$\Gamma \longrightarrow F(\bar{X}_m){:}\mathbf{y} $}
\AxiomC{$\Gamma, F{:}f, X_1{:}x_1, ... ,X{:}x_m \longrightarrow  \mathtt{M}$}
\RightLabel{(app-INST)}
\BinaryInfC{$\Gamma  \longrightarrow \mathtt{M}$}
\end{prooftree}
\end{minipage}
\qquad
\begin{minipage}{0.4\textwidth}
\begin{prooftree}
\AxiomC{$\Gamma , \lambda \tilde{x}_m . Y{:}f \longrightarrow \mathtt{M}$}
\RightLabel{($\lambda$-INST)}
\UnaryInfC{$\Gamma \longrightarrow \mathtt{M}$}
\end{prooftree}
\end{minipage}
\medskip

\qquad
\qquad
\qquad
\qquad
\qquad
\qquad
\begin{minipage}{0.4\textwidth}
\begin{prooftree}
\AxiomC{$\Gamma \longrightarrow  C_{(D_{(k)}/x)} {:} o$}
\AxiomC{$\Gamma , D_{(k)} {:} d \longrightarrow \mathtt{M}$}
\RightLabel{($\lfloor\!\!\lfloor \cdot
\rfloor\!\!\rfloor_o$-INST)}
\BinaryInfC{$\Gamma \longrightarrow \mathtt{M}$}
\end{prooftree}
\end{minipage}

\medskip
\normalsize

{\small
\textit{Remark}. 
(AX) is typically used for an introduction of an assumption into a piece of inference. 
(WR) weakens $\Gamma$ (that entails $\mathtt{M}$) by adding a redundant match $\mathtt{M}'$. 
($\exists$-I) says that if it is true that some object $\mathrm{A}$ is an $\mathrm{F}$, then it is true that $\mathrm{F}$ is non-empty.
($\beta$-EXP) complements the fundamental rule of $\lambda$-calculus, the rule of $\beta$-reduction which captures an application of a function to an argument.
(app-INST) roughly says that if $F(X)$ is $v$-proper,  assumptions that both $F$ and $X$ are $v$-proper are unnecessary in entailing $\mathtt{M}$. 
($\lambda$-INST) says that the assumption that a $\lambda$-abstract is $v$-proper is redundant in entailing any match  $\mathtt{M}$.
($\lfloor\!\!\lfloor \cdot
\rfloor\!\!\rfloor_\tau$-INST) says that since $\Gamma$ entails 
$C_{(D_{(k)}/x)}$  that is $v$-proper, its component part $D_{(k)}$ is $v$-proper; 
$\mathtt{M}$ is entailed by $\Gamma$ even without that assumption.
}

\section{Formulation and derivation of the rule (EG)}
\label{egformulationderivation}

Now we are ready to state (EG) and prove it from the aforementioned rules of $\mathsf{ND_{TT^*}}$.
I thus provide a demonstration for Solution 4 according to which (EG) is not a primitive rule, but a rule derived from the primitive rule ($\exists$-I).

\begin{theorem}
[The rule (EG)]
Let $x /\tau; D / \tau_{(k)}; C, \mathrm{T} /o; \exists^\tau / (\tau \mapsto o) \mapsto o; 
Sub/\langle *^n, *^n, *^n \rangle \mapsto *^n;$\linebreak
$\lfloor\!\!\lfloor \cdot \rfloor\!\!\rfloor_\tau / *^n \mapsto \tau^n; 
0 \leqslant k \leqslant 1$. 
Then,

\begin{prooftree}
\AxiomC{ }  
\RightLabel{$\mathrm{(EG)}$}
\UnaryInfC{$ \Gamma, C_{(D_{(k)}/x)} {:} \mathrm{T}  \longrightarrow \exists^\tau  (\lambda x . C) {:} \mathrm{T}$}
\end{prooftree}

\noindent
is a valid derived rule of $\mathsf{ND_{TT^*}}$.
\end{theorem}

{\small
\textit{Remarks.}
My formulation of (EG) is seemingly similar to Tich\'{y}'s \cite{tichy1986-indiscernibility}. However, he framed it within his STT which does not manipulate constructions or functions from or to constructions; his logic did not (and could not) include constructions of the forms $\ulcorner C \urcorner$ and applications $\lfloor\!\!\lfloor C \rfloor\!\!\rfloor_\tau$. 
His substitution function was metalinguistic, not `intra-logical'; thus, his $C_{(D_{(k)}/x)}$ is in no sense a notational abbreviation of our $ \lfloor\!\!\lfloor Sub(\ulcorner D_{(k)} \urcorner, \ulcorner x \urcorner,\ulcorner C \urcorner) \rfloor\!\!\rfloor_\tau$. 
Another dissimilarity is due to the fact that Tich\'{y}'s rule was adapted to his higher-order logic TIL, which led to unnecessary complications. 
Moreover, Tich\'{y}'s early substitution function has no standard mechanism of avoiding a collision of variables (cf. definition of Sub, point II.iii-iv, for an example of the standard mechanism), which he compensated by supplementary conditions imposed on his rules to preserve their validity.
All differences led Tich\'{y} to the formulation of a metalinguistic proof which differs from my `intra-logical' proof within $\mathsf{ND_{TT^*}}$.
Recall also that the present formulation of (EG) and its proof (below)
differs from the formulation and proof occurring in \cite{raclavsky2021-eg}, for $C_{(D/x)}$ occurring in (EG) is an application $ \lfloor\!\!\lfloor Sub(\ulcorner D_{(k)} \urcorner, \ulcorner x \urcorner,\ulcorner C \urcorner) \rfloor\!\!\rfloor_\tau$, not an immersion (a special form of constructions adjusted from Tich\'y \cite{tichy1988-ffl}).
}

\textit{Proof}. We are going to show that (EG)'s end sequent is derivable using $\mathsf{ND_{TT^*}}$'s primitive rules.
Let $d/\tau$ and the rest as above (similarly below). 

\vspace{-10pt}

\begin{prooftree}
\footnotesize

\AxiomC{}
\RightLabel{(AX)}
\UnaryInfC{$\Gamma , C_{(D_{(k)}/x)} {:} \mathrm{T}  
\longrightarrow C_{(D_{(k)}/x)} {:} \mathrm{T}  
\;\; ^{1.}$}
\RightLabel{(WR)}
\UnaryInfC{$\Gamma , C_{(D_{(k)}/x)} {:} \mathrm{T} , D_{(k)} {:} d  
\longrightarrow C_{(D_{(k)}/x)} {:} \mathrm{T} 
\;\; ^{2.}$}
	
\AxiomC{}
\RightLabel{(AX)}
\UnaryInfC{$\Gamma , D_{(k)} {:} d \longrightarrow D_{(k)} {:} d$}
\RightLabel{(WR)}
\UnaryInfC{$\Gamma , C_{(D_{(k)}/x)} {:} \mathrm{T} , D_{(k)} {:} d  \longrightarrow D_{(k)} {:} d 
\;\; ^{3.}$}

\RightLabel{($\beta$-EXP) [2,3]}
\BinaryInfC{$\Gamma ,  C_{(D_{(k)}/x)} {:} \mathrm{T} , D_{(k)} {:} d  \longrightarrow [\lambda x . C] (D_{(k)}) {:} \mathrm{T} 
$} 

\RightLabel{($\exists$-I)} 
\UnaryInfC{$\Gamma , C_{(D_{(k)}/x)} {:} \mathrm{T} , D_{(k)} {:} d 
\longrightarrow \exists^\tau (\lambda x . C) {:} \mathrm{T}
\;\; ^{4.}$}

\RightLabel{($\lfloor\!\!\lfloor\cdot\rfloor\!\!\rfloor_o$-INST) [1,4]}
\UnaryInfC{$\Gamma , C_{(D_{(k)}/x)} {:} \mathrm{T} \longrightarrow \exists^\tau (\lambda x. C) {:} \mathrm{T} $}

\end{prooftree}

Also the following demonstration is needed for Solution  4.

\begin{theorem}
[The rule $\mathrm{(\exists}$-$\mathrm{I}^\eta)$]
Let $X,x /\tau; F/\tau \mapsto o; o, \mathrm{T} /o; \exists^\tau / (\tau \mapsto o) \mapsto o$.
Then,

\begin{prooftree}
\AxiomC{ }  
\RightLabel{$\mathrm{(\exists}$-$\mathrm{I}^\eta)$}
\UnaryInfC{$ \Gamma, F(X) {:} \mathrm{T}  \longrightarrow \exists^\tau  (\lambda x . F(x)) {:} \mathrm{T}$}
\end{prooftree}

\noindent
is a valid derived rule of $\mathsf{ND_{TT^*}}$.
\end{theorem}

\textit{Proof}. 
To simplify the following tree (an adaptation of the proof stated in \cite{raclavsky2021-eg}), let 

\begin{prooftree}
\scriptsize
\AxiomC{}
\LeftLabel{$\mathtt{D}:= $}
\RightLabel{(AX)}
\UnaryInfC{$\Gamma , F(X) {:} \mathrm{T}  \longrightarrow F(X) {:} \mathrm{T}  \;\;^{4.}$}
\end{prooftree}

\vspace{-10pt}

\begin{prooftree}
\scriptsize

\AxiomC{}
\RightLabel{(AX)}
\UnaryInfC{$\Gamma , F(x)_{(X/x)} {:} \mathrm{T} , 
F{:}f, X{:}x \longrightarrow F(x)_{(X/x)}  {:} \mathrm{T} \;\;^{1.} $}

\AxiomC{}
\RightLabel{(AX)}
\UnaryInfC{$\Gamma , F(x)_{(X/x)} {:} \mathrm{T}, F{:}f, X{:}x  \longrightarrow 
X{:}x \;\;^{2.}$}

\RightLabel{($\beta$-EXP) [1,2]}
\BinaryInfC{$\Gamma , F(x)_{(X/x)} {:} \mathrm{T} , F{:}f, X{:}x\longrightarrow 
[\lambda x . F(x) ] (X) {:} \mathrm{T} $} 

\RightLabel{Def of Sub}
\UnaryInfC{$\Gamma , F(X) {:} \mathrm{T} , F{:}f, X{:}x\longrightarrow 
[\lambda x . F(x) ] (X) {:} \mathrm{T} $} 

\RightLabel{($\exists$-I)}
\UnaryInfC{$\Gamma , F(X) {:} \mathrm{T} , F{:}f, X{:}x \longrightarrow 
\exists^\tau (\lambda x . F(x)) {:} \mathrm{T} \; \; ^{3.}$} 

\AxiomC{$\mathtt{D}$}

\RightLabel{(app-INST) [3,4]}
\BinaryInfC{$\Gamma , F(X) {:} \mathrm{T} \longrightarrow 
\exists^\tau (\lambda x . F(x)) {:} \mathrm{T} $}
\end{prooftree}

As mentioned above, ($\exists$-I$^\eta$) appears to be derivable by an application of the rule of $\eta$-conversion which says that 
the `functions' $F$ and $\lambda x. F(x)$ are extensionally identical.
In $\mathsf{ND_{TT^*}}$, 
the rule of $\eta$-contraction requires $F$ to be $v$-proper in order to prevent failure because of partiality (see e.g. Raclavsk\'y \cite{raclavsky2010-partiality}, with Kuchy\v{n}ka and Pezlar \cite{raclavsky-kuchynka-pezlar2015}).

\begin{prooftree}
\small
\AxiomC{$\Gamma \longrightarrow \lambda x. F(x) {:}f$}
\AxiomC{$\Gamma \longrightarrow F{:}\mathbf{g}$}
\RightLabel{($\eta$-CON)}
\BinaryInfC{$\Gamma \longrightarrow F{:}f$}
\end{prooftree}

\begin{prooftree}
\small
\AxiomC{$\Gamma , D_1{:}\mathbf{x}_1 \longrightarrow 
D_2{:}\mathbf{x}_2$}
\AxiomC{$\Gamma , D_2{:}\mathbf{x}_2 \longrightarrow 
D_1{:}\mathbf{x}_1$}
\RightLabel{(RT)}
\BinaryInfC{$\Gamma , C_{(D_2/x_2)}{:}o \longrightarrow  C_{(D_1/x_1)}{:}o$}
\end{prooftree}

To simplify the following tree, let 

\begin{prooftree}
\footnotesize
\AxiomC{}
\LeftLabel{$\mathtt{D}:=$}
\RightLabel{(AX)}
\UnaryInfC{$\Gamma, \lambda x. F(x){:}f,  F(X){:}\mathrm{T} \longrightarrow F(X){:}\mathrm{T}$ \,$^{5.}$}
\end{prooftree}

\begin{prooftree}
\footnotesize

\AxiomC{}
\RightLabel{(AX)}
\UnaryInfC{$\Gamma, F{:}f , \lambda x . F(x){:}f \longrightarrow \lambda x. F(x) {:}f$\;$^{1.}$}

\AxiomC{}
\RightLabel{(AX)}
\UnaryInfC{$\Gamma, \lambda x . F(x){:}f , F{:}f  \longrightarrow F{:}f$ \;$^{2.}$}

\RightLabel{($\eta$-CON) [1,2]}
\BinaryInfC{$\Gamma, F{:}f , \lambda x . F(x){:}f \longrightarrow F{:}f$ \;$^{3.}$}


\RightLabel{(RT) [1,3]}
\UnaryInfC{$\Gamma, F{:}f, \lambda x. F(x){:}f, f(X)_{(F/f)}{:}\mathrm{T} \longrightarrow f(X)_{(\lambda x. F(x)/f)}{:}\mathrm{T} $}

\RightLabel{Def of Sub}
\UnaryInfC{$\Gamma, F{:}f, \lambda x. F(x){:}f, F(X){:}\mathrm{T} \longrightarrow [\lambda x. F(x)](X) {:}\mathrm{T}$}

\RightLabel{(WR)}
\UnaryInfC{$\Gamma, F{:}f, X{:}x, \lambda x. F(x){:}f, F(X){:}\mathrm{T} \longrightarrow [\lambda x. F(x)](X) {:}\mathrm{T} $ \,$^{4.}$}

\AxiomC{$\mathtt{D}$}

\RightLabel{(app-INST) [4,5]}
\BinaryInfC{$\Gamma, \lambda x. F(x){:}f, F(X){:}\mathrm{T} \longrightarrow [\lambda x. F(x)](X) {:}\mathrm{T} $}

\RightLabel{($\lambda$-INST)}
\UnaryInfC{$\Gamma, F(X){:}\mathrm{T} \longrightarrow [\lambda x. F(x)](X) {:}\mathrm{T} $}

\RightLabel{($\exists$-I)}
\UnaryInfC{$\Gamma, F(X){:}\mathrm{T} \longrightarrow \exists^\tau (\lambda x . F (x)) {:}\mathrm{T} $}

\end{prooftree}

\noindent
However, once we recall that in $\mathsf{ND_{TT^*}}$ the antecedents of sequents are not lists but sets, we notice that sequents 1 and 2 are identical. In consequence, sequent 2 becomes dispensable, the use of ($\eta$-CON) in deriving ($\exists$-I$^\eta$) is thus redundant.

\section{Quantification into}
\label{quantificationinto}

In this section, I first describe how to demonstrate that (EG) justifies an argument (Sec. \ref{justificationarguments}). 
Then, I explain how 
Sub's definition prevents such justification in cases of arguments that are only seemingly justified by (EG) (Sec. \ref{subinapplicability}).

As mentioned in the introductory section,
the present section overlaps with the corresponding selection from \cite{raclavsky2021-eg}. The aim of that is twofold: to
show that even the present approach delivers the right desiderata and allow a possible comparison of  the present approach with that in \cite{raclavsky2021-eg}.

\subsection{Justification of an argument by (EG)}
\label{justificationarguments}

An argument $A$ is called \textit{justified w.r.t. the set of rules $R$}, if it is possible to derive -- using members of $R$ -- the sequent 
of the form 
$$\Gamma, P_1{:}\mathrm{T}, ..., P_m{:}\mathrm{T}\longrightarrow P{:}\mathrm{T}$$ where $\bar{P}_m$ are (formalizations of) $A$'s premisses
and $P$ is a (formalization of) $A$'s conclusion.
The crucial non-primitive rules of $\mathsf{ND_{TT^*}}$ used in such a derivation are indicated 
on the right-hand side of $A$'s $\rule{1.0cm}{0.4pt}$. 

Here is a straightforward proof demonstrating the justification of the argument 
\medskip

\begin{minipage}{.2\textwidth}
\hfill $A^E$ \;\;
\end{minipage}
 \begin{minipage}{.4\textwidth}
\begin{prooftree}
 \small
\AxiomC{\textsl{$3{\div}1$ is odd.}}
\RightLabel{(EG)}
\UnaryInfC{\textsl{Some number $n$ is such that $3{\div}n$ is odd.}}
\end{prooftree}
\end{minipage}

\medskip
\noindent
w.r.t. $R=\{$(EG)$\}$. Let $n, 0,1,3/\nu; \div/\langle \nu, \nu\rangle \mapsto \nu; \mathrm{Odd}/\nu \mapsto o$.

\begin{prooftree}
\small
\AxiomC{}
\RightLabel{(AX)}
\UnaryInfC{$\Gamma , \mathrm{Odd} (3{\div }n)_{(1/n)}  {:} \mathrm{T} 
\longrightarrow  \mathrm{Odd} (3{\div }n)_{(1/n)}   {:} \mathrm{T}$}
\RightLabel{Def of Sub}
\UnaryInfC{$\Gamma , \mathrm{Odd} (3{\div }1)  {:} \mathrm{T} 
\longrightarrow  \mathrm{Odd} (3{\div }n) _{(1/n)} {:} \mathrm{T}$}
\RightLabel{(EG)}
\UnaryInfC{$\Gamma , \mathrm{Odd} (3{\div }1)   {:} \mathrm{T} 
\longrightarrow \exists^\nu (\lambda  n . \mathrm{Odd} (3{\div } n) ) {:} \mathrm{T}$}
\end{prooftree}

Arguments involving I- or H-contexts and existential generalization 
can be controlled in a similar way (in their \textit{de re} readings, of course).

\subsection{Inapplicability of (EG) in case of \textit{de dicto} readings of I- and H-contexts}
\label{subinapplicability}

The present section documents that our Sub is satisfactorily defined for handling quantification to I- and H-contexts, as suggested in Solution 6. I explain why some arguments involving existential generalisation are not justified by $R=\{$(EG)$\}$, while it accords with our intuitions. We will see that the impossibility is encoded in our definition of Sub. 

We will show two principal cases: (a) modality \textit{de dicto} (a frequent case of I-contexts) and (b) (\textit{de dicto}) H-contexts. Both are known for impossibility to quantify into them. 
I begin with the case of H-contexts, for it provides an easy illustration of the key point.

\subsubsection{The case of H-contexts}

When demonstrating a justification of $A^H_1$ or $A^H_2$ 
w.r.t. $R=\{(EG) \}$, one must successfully apply Sub's definition to the \textit{Sub-form} (as I will call it) of the relevant argument's premiss. The Sub-form exhibits a substitution into $C$, while the result of the substitution is the premiss.
$C$ also occurs in the argument's conclusion $\exists^\tau (\lambda x. C)$. 

In our present case with $A^H_1$ we have the following Sub-form. Let $c^1/*^1$ be a variable for $1$st-order constructions such as $3{\div}0$;
$\mathrm{Improp} / *^1 \mapsto o$ and the rest as above.
Since it is $3{\div}0$ itself (not its non-existent value) which is undefined, $3{\div}0$ must be `delivered' by $\ulcorner 3{\div}0  \urcorner$ which is the simplest construction of the construction $3{\div}0$:

\begin{center}
Sub-$A^H_1$
\qquad
$\lfloor\!\!\lfloor{Sub}^2
(\ulcorner \ulcorner 3{\div}0 \urcorner \urcorner , \ulcorner c^1 \urcorner ,
\ulcorner \mathrm{Improp}(c^1) \urcorner ) \rfloor\!\!\rfloor_o
$
\end{center}

On the other hand, a suitable Sub-form does not exist in the case with $A^H_2$. Its premiss' alleged Sub-form
$$
\lfloor\!\!\lfloor{Sub}^2
(\ulcorner 0 \urcorner , \ulcorner n \urcorner ,
\ulcorner \mathrm{Improp}(\ulcorner 3{\div}n\urcorner) \urcorner ) \rfloor\!\!\rfloor_o
$$
is not $v$-congruent to $A^H_2$'s premiss. 
Because
Sub$^2$ cannot substitute a construction for a bound variable, namely for the variable $n$ in the $2$nd-order construction 
$C:= \mathrm{Improp}(\ulcorner 3{\div}n\urcorner)$.\footnote{Since every construction of order $n$ is ranked also as of order $n+1$, see the definition of type above, both $n$ and $3{\div}0$ have an order suitable for application of Sub$^2$.} That $n$ is not free in $C$, for it occurs in the scope of  $\ulcorner \cdot \urcorner$, i.e. within an `opaque' context. 
$A^H_2$ is thus not justified by (EG).

The just presented reading of $A^H_2$ might perhaps be called `\textit{de dicto}' and lead us to seeking its `\textit{de re}' counterpart, $A^H_{2'}$, that says that a number $\mathrm{N}$ is such that 
the construction which is the result of its substitution in $3{\div}n$ is improper.

Let me explain that the appropriate $A^H_{2'}$'s Sub-form is 

\begin{center}
Sub-$A^H_{2'}$
\qquad
$\lfloor\!\!\lfloor{Sub}^2
(\ulcorner 0 \urcorner , \ulcorner n \urcorner ,
\ulcorner 
\mathrm{Improp} ( {Sub}^1  (\ulcorner (n) \urcorner, \ulcorner n \urcorner ,
\ulcorner 3{\div}n\urcorner)  ) \urcorner )
\rfloor\!\!\rfloor_o$
\end{center}

\noindent
Sub-$A^H_{2'}$ exhibits a quantification into the construction that applies the characteristic function Improper to the construction $v$-constructed by $C:=$
${Sub}^1 (\ulcorner ( n ) \urcorner , \ulcorner n \urcorner , \ulcorner 3{\div}n\urcorner)$. In $C$, $n$ is free since it has a free occurrence in $\ulcorner ( n ) \urcorner $, which is a construction 
employing the function denoted by ``$\ulcorner ( \cdot ) \urcorner $'' (it is an adaptation of Tich\'y's \cite{tichy1988-ffl} trivialization function). The partial function $\ulcorner ( n ) \urcorner $ maps the number $\mathrm{N}$ $v$-constructed by $n$ to the unique $1$st-order construction $\mathbf{N}$  to which Sub$^1$ is applied. 
The reason for this move: 
Sub$^1$ could not be applied to a number, only to a construction of the number; each number is $v$-constructed by infinitely many constructions, but one needs	 to choose a unique one; 
the construction of the form $\ulcorner (X) \urcorner $ is then the obvious choice since 
$\ulcorner (X) \urcorner $ delivers 
$\ulcorner \mathrm{X} \urcorner $ that 
serves as a proper name of the object $\mathrm{X}$ $v$-constructed by $X$ (if $\mathrm{X}$ is not a construction, but an object, $\ulcorner X \urcorner $ reads $\mathbf{X} $).

\subsubsection{The case of I-contexts}

Even in the case of sentences involving modalities that create I-contexts we obtain results that match our expectations. 

First, consider $A^I_1$. Let $D/\omega \mapsto \iota; F/ \omega \mapsto (o \mapsto \iota) ; \forall^\omega / (\omega \mapsto o) \mapsto o $. $A^I_1$'s premiss is\linebreak
 $\forall^{\omega}( \lambda w' . [ F(w')] (D(w))  )$; its Sub-form is:

\begin{center}
Sub-$A^I_1$ \qquad
$\lfloor\!\!\lfloor Sub^1 (\ulcorner D(w) \urcorner , \ulcorner x \urcorner , \ulcorner \forall^{\omega}( \lambda w' . [ F(w')] (x)  ) \urcorner ) \rfloor\!\!\rfloor_o$
\end{center}

\noindent
Due to Sub's definition, the variable $w$ of $D(w)$ does not become $\lambda$-bound when carrying out the substitution.

On the other hand, no such suitable Sub-form exists in the case  of $A^I_2$. 
The alleged candidate 
$$\lfloor\!\!\lfloor Sub^1 (\ulcorner D(w') \urcorner , \ulcorner x \urcorner , \ulcorner \forall^{\omega}( \lambda w' . [ F(w')] (x) )  \urcorner ) \rfloor\!\!\rfloor_o$$
is not $v$-congruent to $A^I_2$'s premiss $\forall^{\omega}( \lambda w' . [ F(w')] (D(w'))  )$, for 
Sub prevents $\lambda$-binding of its variables, cf. definition of Sub, point II.iv. And so after substitution, $D$ remains applied to a free variable (similarly as in the case with $A^I_1$). $A^I_2$ is thus not justified by $R=\{$(EG)$\}$.

\section{Conclusions}
\label{conclusion}

The present paper utilises the 
often neglected (cf. \cite{uzquiano2020-sep}) 
approach of STT to $\exists$, according to which $\exists$ is an operator that denotes the function that maps all non-empty sets to $\mathtt{T}$ and the empty-set(s) to $\mathtt{F}$. 
This definition holds even in partial logics such as $\mathsf{TT^*}$ which contain STT in its core. 
Moreover, the approach leads to easy solutions to the various puzzles related to the rule (EG), see our Solutions 1 -- 6 and also Facts 1 -- 10 about $\exists$ and (EG).
(The usability of ideas based on this within FOL, which does not involve quantification over sets, requires that we understand its formula $\exists x F(x)$ as indirectly saying that the denotation of the predicate $F$ is non-empty.)

The present paper
provides also an essential revision of my previous paper \cite{raclavsky2021-eg}. It consists in an entire omission of the constructions (called immersions) of the form $\lfloor\!\!\lfloor C \rfloor \!\!\rfloor _\tau$ from $\mathsf{TT^*}$. 
\cite{raclavsky2021-eg}'s explication of  
$C_{(D/x)}$ as a certain immersion $\lfloor\!\!\lfloor C \rfloor \!\!\rfloor _\tau$ can be neglected.
In this paper, the explicit substitution operator $C_{(D/x)}$ is rather identified with an application of the form $F(A)$, where $F$ is a construction of the partial function $\lfloor\!\!\lfloor \cdot \rfloor \!\!\rfloor _\tau$ that 
maps constructions to the objects (if any) $v$-constructed by them; and $A$ is an application $Sub(\ulcorner D\urcorner, \ulcorner x \urcorner, \ulcorner C \urcorner)$
which applies the substitution function to the triple $\langle D,x,C\rangle$ and returns $C'$, which is the result of the substitution of $D$ for $x$ in $C$. 
This novel approach required novel proofs of rules such as (EG). 
Such `objectual' explanation of $C_{(D/x)}$ is an important part of the overall $\mathsf{TT^*}$ approach, since its rules need no external `metalinguistic' specification, they are fully specifiable inside the system.

\subsection*{Acknowledgement}
I am grateful to the reviewers for their remarks that helped to improve the present paper. The work on this paper was supported by the grant of the Czech Science Foundation (GA\v{C}R) registration no. GA19-12420S, ``Hyperintensional Meaning, Type Theory and Logical Deduction''.

\section*{Appendix}
\label{sec:appendix}

The following provides rules defining (in a general way) the \textit{evaluation function} $[\![ C]\!]^{\mathscr{M},v}$, which maps constructions $C$
to the objects (if any, cf. $\_$) constructed by the constructions 
w.r.t. an assignment $v$ and a model $\mathscr{M}$.
Let $X$ be an object or a construction, $C, \bar{C}_m$ constructions, $c$ a variable for constructions, and $\mathrm{C},\bar{\mathrm{C}}_m$ (mind the upright font) objects $v$-constructed by $C, \bar{C}_m$, respectively:
\medskip 

\begin{center}

\begin{tabular}{p{140pt}lp{260pt}}

$[\![ x_i^\tau]\!]^{\mathscr{M},v}$ & 
$=$ &
the $i$th-member $X_i$ of the sequence $sq^\tau$ of $\tau$-objects, 
where $sq^\tau$ belongs to $v$
\\
$[\![ \mathit{C}]\!]^{\mathscr{M},v}$ & 
$=$ &
$\mathscr{I}(\mathit{C})$ where $\mathit{C}$ is a constant-as-construction
\\
\end{tabular}

\medskip

\begin{tabular}{p{140pt}lp{260pt}}
$[\![ \ulcorner C \urcorner]\!]^{\mathscr{M},v}$ & $\, =$ &
$C$
\end{tabular}

\medskip

\begin{tabular}{l}
$
[\![ C(\tilde{C}_m)]\!]^{\mathscr{M},v} \quad\, 
\qquad\qquad\qquad\quad\;\
=  \quad
\!\begin{cases} 
\mathrm{C}(\bar{\mathrm{C}}_m)
 &
\mbox{if } 
[\![ C]\!]^{\mathscr{M},v}  = \mathrm{C} \in \mathscr{D}_{\langle \bar{\tau}_m \rangle \mapsto \tau}, [\![ C_1 ]\!]^{\mathscr{M},v}  =  \mathrm{C}_1 \in \mathscr{D}_{\tau_1}, \\
&
..., [\![ C_m ]\!]^{\mathscr{M},v}  =  \mathrm{C}_m \in \mathscr{D}_{\tau_m} \mathrm{\,and\,}
\exists x (x = \mathrm{C}(\bar{\mathrm{C}}_m)) 
\\
 & 
\mbox{otherwise} 
\end{cases}  
$
\end{tabular}

\medskip

\begin{tabular}{p{140pt}lp{260pt}}
$[\![ \lambda \tilde{x}_m.C]\!]^{\mathscr{M},v}$ & = &
the function $f \in \mathscr{D}_{\langle \bar{t}_m \rangle  \mapsto \tau}$ that takes each $[\![ C ]\!]^{\mathscr{M},v^{(')}} \in \mathscr{D}_\tau$
at the respective argument
$\langle [\![ x_1]\!] ^{\mathscr{M},v^{(')}},...,[\![ x_m ]\!]^{\mathscr{M},v^{(')}} \rangle$
where $v'$ is like $v$ except for what it assigns to $\bar{x}_m$ and for each $1 \leqslant i \leqslant m$, 
$[\![ x_i ]\!]^{\mathscr{M},v^{(')}} \in \mathscr{D}_{\tau_i}$ 
\\
\end{tabular}

\end{center}

\bibliographystyle{eptcs}
\bibliography{biblio}

\end{document}